\documentclass[12pt]{article}
\usepackage{amscd,amssymb,amsmath,latexsym,enumerate}
\usepackage[mathscr]{euscript}
\usepackage{epsfig}
\usepackage{fancybox}
\usepackage{verbatim}
\usepackage{stmaryrd}

\usepackage[small,nohug]{diagrams}
\diagramstyle[labelstyle=\scriptstyle]

\usepackage{color}

\title{Topological boundary invariants \\ for Floquet systems and quantum walks}

\author{Christian Sadel$^1$, Hermann Schulz-Baldes$^2$
\\
\\
{\small $^1$ Facultad de Matem\'aticas, Pontificia Universidad Cat\`olica de Chile, Chile}
\\
{\small $^2$ Department Mathematik, Friedrich-Alexander-Universit\"at Erlangen-N\"urnberg, Germany}
}

\date{ }

\newtheorem{theo}{Theorem}
\newtheorem{defini}{Definition}
\newtheorem{proposi}{Proposition}

\newtheorem{coro}{Corollary}

\newcommand{\CM}{{\mathbb C}}
\newcommand{\NM}{{\mathbb N}}
\newcommand{\RM}{{\mathbb R}}
\newcommand{\SM}{{\mathbb S}}

\newcommand{\ZM}{{\mathbb Z}}
\newcommand{\PM}{{\mathbb P}}

\newcommand{\Aa}{{\cal A}}
\newcommand{\Ee}{{\cal E}}

\newcommand{\BB}{{\bf B}}
\newcommand{\EE}{{\bf E}}
\newcommand{\Bb}{{\cal B}}

\newcommand{\Ss}{{\cal S}}

\newcommand{\Tt}{{\cal T}}

\newcommand{\one}{{\bf 1}}

\newcommand{\TR}{{\rm Tr}} 
\newcommand{\Tr}{\mbox{\rm Tr}}

\newcommand{\Ch}{{\rm Ch}} 
\newcommand{\Ind}{{\rm Ind}} 
 
\newcommand{\Exp}{{\rm Exp}}

\newcommand{\diag}{{\rm diag}}

\begin{document}

\maketitle

\begin{abstract}
A Floquet systems is a periodically driven quantum system. It can be described by a Floquet operator. If this unitary operator has a gap in the spectrum, then one can define associated topological bulk invariants which can either only depend on the bands of the Floquet operator or also on the time as a variable. It is shown how a $K$-theoretic result combined with the bulk-boundary correspondence leads to edge invariants for the half-space Floquet operators. These results also apply to topological quantum walks.
\end{abstract}

\section{Introduction}

In the theory of fermionic topological insulators one associates non-trivial topological invariants to the bands of the possibly disordered one-particle Hamiltonians  \cite{RSFL,PS}. For a periodically driven quantum system, it is convenient to study the associated Floquet operator and its quasi-energy spectrum. For its bands, one can readily apply the standard definitions of topological invariants (even in presence of disorder), but as stressed in \cite{RLBL} and further developed in \cite{Fru} it is also possible to define new topological invariants which involve the time as a supplementary parameter. For two-dimensional periodically driven systems, it was then argued in \cite{RLBL} that these new invariants can lead to topological edge states for the half-space Floquet operator, even if the bands themselves have vanishing bulk invariants. These new states were called anomalous edge states \cite{RLBL}. A similar phenomenon also appears for one-dimensional driven systems if a chiral symmetry is present \cite{ATD,Fru}.  In this paper, the claims about anomalous edge states from \cite{RLBL} are rigorously proved also for disordered time-periodic Hamiltonians and in arbitrary space dimension.

\vspace{.2cm}

The proof is based on the bulk-boundary correspondence for topological insulators \cite{PS} combined with a key mathematical result on connecting maps in $K$-theory. It is formulated in Theorem~\ref{theo-ChBBC3} for topological Floquet systems, but the central part of the proof actually leads to the following abstract statement that we believe to be of independent interest. In its formulation, standard notions of $K$-theory as described in \cite{RLL,WO,PS} will be freely used.

\begin{theo}
\label{theo-ThetaInd} 
Let $0\to\Ee\to\Bb\to\Aa\to 0$ be an exact sequence of C$^*$-algebras with $\Bb$ and $\Aa$ unital. Suppose given a differentiable path $(0,2\pi)\mapsto U(t)\in\Aa$ specifying a class in the $K_1$-group $K_1(S\Aa)$ of the suspension of $\Aa$. Under the index map $\Ind:K_1(S\Aa)\to K_0(S\Ee)$ and inverse suspension map $\Theta^{-1}:K_0(S\Ee)\to K_1(\Ee)$, one has
$$
\Theta^{-1}(\Ind ([(0,2\pi)\mapsto U(t)]_1))
\;=\;
[\widehat{U}(2\pi)]_1
\;,
$$
where $\widehat{U}(2\pi)-\one\in\Ee$ is the end point of the initial value problem
$$
\imath\,\partial_t \widehat{U}(t)
\;=\;
\widehat{H}(t)\,\widehat{U}(t)
\;,
\qquad
\widehat{U}(0)\,=\,\one
\;,
$$
associated to a self-adjoint lift $\widehat{H}(t)\in\Bb$ of $H(t)=-\imath\,U(t)\partial_tU(t)^*\in\Aa$.
\end{theo}

Let us stress that the initial value problem is solved in $\Bb$ so that the path $(0,2\pi)\mapsto \widehat{U}(t)$ lies in $\Bb$, but it is shown that the end point $\widehat{U}(2\pi)-\one$ lies in the ideal $\Ee$. Furthermore, the identity $\Theta^{-1}\circ\Ind=\Exp\circ\beta^{-1}$ allows to rewrite the result in terms of the exponential map and Bott map. The proof of this result will be given in Section~\ref{sec-BBC} after the proof of Theorem~\ref{theo-ChBBC3}.

\vspace{.2cm}

Quantum walks is another field of research where gapped unitary operators over a lattice Hilbert space appear naturally \cite{Kit,PC,DFT,ABJ}. The archetypical two-dimensional quantum walk is the Chalker-Coddington model \cite{CC}. When restricted to a half-space, it has topological edge states \cite{HC}. Their topological origin was uncovered only very recently \cite{DFT}. In fact, it was shown in \cite{DFT} that any quantum walk with topologically trivial steps can be understood as a periodically driven system with suitably chosen Hamiltonian. The Floquet operator of this system then coincides with the unitary evolution operator of the quantum walk. Hence the bulk-boundary correspondence proved below can also be applied to quantum walks. Interestingly, for the Chalker-Coddington model all bands have vanishing standard invariants (namely, Chern numbers) and the edge states found in \cite{HC} are actually anomalous.

\vspace{.2cm}

Below we will freely use  the description of covariant operator families by means of a crossed product C$^*$-algebra as well as the non-commutative analysis tools thereon. A brief summary of the notations is included for the convenience of the reader in Appendix~\ref{app-alg}, but it may be necessary to go back to the detailed treatment given in \cite{Bel} or \cite{PS}.
 
\vspace{.3cm}

\noindent {\bf Acknowledgements:} The authors thank Rafeal Tiedra for discussions on quantum walks, and the Instituto de Matematicas (UNAM), Cuernavaca, for its hospitality during a visit in July 2017 when this work was written. We also thank an unknown referee for a careful reading and constructive comments. An independent contribution on similar matters by Graf and Tauber \cite{GT} appeared on the {\tt arXiv} while preparing the final version of this manuscript. It also proves bulk-boundary correspondence for Floquet systems, but is restricted to two-dimensional models. It provides a more ad hoc functional analytic treatment, while here the general theory of bulk-boundary correspondence of topological insulators from \cite{PS} is combined with a  $K$-theoretic fact (Theorem~\ref{theo-ChBBC3}) not contained in \cite{GT}. Before preparing the final revision of this work, another related draft on edge states in the Chalker-Coddington model was posted by Asch, Bourget and Joye  \cite{ABJ2}. This research was partly supported by the Chilean grant FONDECYT Regular 1161651 and the DFG.

\section{Bulk Floquet operators and their invariants}
\label{sec-BulkInv}

Let $t\mapsto H(t)=H(t)^*\in\Aa_d$ be a $2\pi$-periodic family of covariant Hamiltonians on $\ell^2(\ZM^d,\CM^L)$, see Appendix~\ref{app-alg} for the definition of such families and the C$^*$-algebra $\Aa_d$. For the following, continuity properties of this path in $t$ are  {\it not} of importance. The associated time-evolution is then a differentiable path of unitaries $t\mapsto U(t)\in\Aa_d$ satisfying
\begin{equation}
\label{eq-TimeEvol}
\imath\,\partial_t U(t)\;=\;H(t)\,U(t)
\;,
\qquad
U(0)\;=\;\one
\;.
\end{equation}
The evolution $U(2\pi)$ over one time period $2\pi$ is then called the Floquet operator and is denoted by $U=U(2\pi)\in\Aa_d$. Its (almost sure) spectrum on $\SM^1$ is called the quasi-energy spectrum. 

\vspace{.2cm}

The Floquet operator $U$ itself is topologically trivial in the sense that $[U]_1=0$ in $K_1(\Aa_d)$ because the solution of \eqref{eq-TimeEvol} provides a path connecting it to the identity. If, however, there are gaps in the quasi-energy spectrum $\sigma(U)$ of $U$, there are interesting topological invariants \cite{RLBL,Fru,DFT}. First let us suppose that $U$ has two gaps $e^{\imath\theta},e^{\imath\theta'}\not\in\sigma(U)$ with $0\leq \theta<\theta'<2\pi$. Then there is a spectral projection $P_{[\theta,\theta']}\in\Aa_d$ of $U$ on the band with quasienergies between $e^{\imath\theta}$ and $e^{\imath\theta'}$. This projection specifies a class $[P_{[\theta,\theta']}]_0$ in $K_0(\Aa_d)$ with even Chern numbers $\Ch_I (P_{[\theta,\theta']})$ defined next. In the notations of \cite{PS}, $I \subset \{1,\ldots,d\}$ is here a subset of even cardinality $n=|I|$ and, for any differentiable projection $P\in\Aa_d$,
\begin{equation}
\label{EvenBulkChernNumbers}
\Ch_I (P) \;=\;  
\frac{(2\imath \pi)^{\frac{n}{2}}}{\frac{n}{2}!}
\;
\sum_{\rho\in \Ss_{I}}(-1)^{\rho}
\;\Tt \Big (P\,\nabla_{\rho_1}P\cdots \nabla_{\rho_n}P \Big)
\;,
\end{equation}
where $\Ss_I$ denotes the $n!$ permutations on $I$ with signature $(-1)^\rho$, $\Tt$ the trace per unit volume and $\nabla_j$ the non-commutative derivatives, {\it cf.} Appendix~\ref{app-alg}. The best-known even Chern number is $\Ch_{\{1,2\}}(P)$ which plays a prominent role in the theory of the integer quantum Hall effect \cite{PS}. The even bulk invariant $\Ch_I (P_{[\theta,\theta']})$ only depends on the Floquet operator $U$, and not on the entire time evolution. It was shown in \cite{RLBL,Fru} that there are further bulk invariants which {\em do} depend on the whole path $t\in[0,2\pi]\mapsto U(t)$. For their definition, suppose that $e^{\imath\theta}\not\in\sigma(U)$ for $\theta\in[0,2\pi)$ and introduce an effective Hamiltonian $h_\theta\in\Aa_d$ by
$$
h_\theta\;=\;-\,\frac{1}{2\pi\imath}\;\log_\theta(U)
\;.
$$
Here $\log_\theta$ denotes the natural logarithm with branch cut along the line $r\in[ 0,\infty)\mapsto re^{\imath\theta}$, namely $\log_\theta(e^{\imath\theta})=\imath(\theta-2\pi)$ and $\lim_{\epsilon\downarrow0}\log_\theta(e^{\imath(\theta-\epsilon)})=\imath \theta$. The spectrum of $h_\theta$ lies in $(-\frac{\theta}{2\pi},1-\frac{\theta}{2\pi})$. By construction, one has $U=e^{-2\pi\imath h_\theta}$. Then set
\begin{equation}
\label{eq-HamTheta}
H_\theta(t)
\;=\;
\left\{
\begin{array}{cc}
2\,H(2t)\;,
& 
t\in[0,\pi]\;,
\\
-\,2\,h_\theta\;,
&
t\in(\pi,2\pi]
\;.
\end{array}
\right.
\end{equation}
Now the periodized time evolution $V_\theta$  is introduced by
\begin{equation}
\label{eq-PeriTimeEvol}
\imath\,\partial_t V_\theta(t)\;=\;H_\theta(t)\,V_\theta(t)
\;,
\qquad
V_\theta(0)\;=\;\one
\;.
\end{equation}
The terminology reflects that
$$
V_\theta(0)\;=\;V_\theta(2\pi)\;=\;\one
\;.
$$
Let us note that the loop $t\in[0,2\pi]\mapsto V_\theta(t)\in\Aa_d$ is homotopic to $t\in[0,2\pi]\mapsto U(t)\,e^{\imath t h_\theta }$ (see also \cite{CDFGT}). These loops specify a class $[V_\theta]_1=[t\in(0,2\pi)\mapsto V_\theta(t)]_1$ in $K_1(S\Aa_d)$ where $S\Aa_d=C_0((0,2\pi),\Aa_d)$ denotes the suspension of $\Aa_d$. More precisely, as $S\Aa_d$ is not unital, and one has the class of the unitary $(t\in(0,2\pi)\mapsto (V_\theta(t)-\one,1))$ in the unitization $(S\Aa_d)^+$.  From this class, one can extract odd Chern numbers $\Ch_J (V_\theta) $, again by using the formulas from \cite{PS}. Let us consider the time to be the $0$-th coordinate and set $\nabla_0=\partial_t$. For an index set $J \subset \{0,1,\ldots,d\}$ of odd cardinality $m=|J|$ and containing $0$, the odd Chern numbers are then defined by ({\it cf.} Sections~2.3.1 and 5.4 in \cite{PS})
\begin{equation}
\label{OddBulkChernNumbers}
\Ch_J (V)  \;=\;   
\frac{\imath\,(\imath \pi)^\frac{m-1}{2}}{m!!}
\sum_{\rho\in \Ss_{J}}(-1)^{\rho}
\;\Tt^s \Big ( (V^* - \one) \nabla_{\rho_1}V
(\nabla_{\rho_2}V^* \nabla_{\rho_3}V) \cdots (\nabla_{\rho_{m-1}}V^*\nabla_{\rho_m}V) \Big)
\;,
\end{equation}
where the suspended trace per unit volume $\Tt^s=\int^{2\pi}_0\frac{dt}{2\pi}\,\Tt$ contains a normalized integral over time, see also \cite{PS}. Note that for $m=1$ this reduces to the winding number, and for $m=3$ to the higher winding number used in \cite{RLBL}. Odd Chern numbers can also be defined for the bands of the Floquet operator when a chiral symmetry is present \cite{Fru}, but this will not be further analyzed here. Now all these invariants are not independent. In fact, as already pointed out in \cite{RLBL,Fru}, the odd Chern numbers $\Ch_J (V_\theta) $ associated to different gaps are connected by the following formulas.

\begin{proposi}
\label{prop-WindChLink}
Let $I\subset\{1,\ldots,d\}$ be an index set of even cardinality. For $e^{\imath\theta},e^{\imath\theta'}\not\in\sigma(U)$ in two gaps of the quasi-energy spectrum  with $0\leq \theta<\theta'<2\pi$, one has
$$
\Ch_{\{0\}\cup I} (V_{\theta'})\,-\,\Ch_{\{0\}\cup I} (V_{\theta})
\;=\;
\Ch_I(P_{[\theta,\theta']})
\;.
$$ 
\end{proposi}

\noindent {\bf Proof.}
This is rooted in a $K$-theoretic fact, namely the Bott map $\beta:K_0(\Aa_d)\to K_1(S\Aa_d)$ satisfies
$$
\beta([P_{[\theta,\theta']}]_0)
\;=\;
[V_{\theta}]_1\,-\,[V_{\theta'}]_1
\;.
$$
Indeed, using $h_{\theta}-h_{\theta'}= P_{[\theta,\theta']}$,
%
$$
[V_{\theta}]_1\,-\,[V_{\theta'}]_1
\;=\;
[V_{\theta'}^*V_{\theta}]_1
\;=\;
[e^{-\imath t (h_{\theta'}-h_\theta)}]_1
\;=\;
[e^{\imath t P_{[\theta,\theta']}}]_1
\;=\;
[(\one-P_{[\theta,\theta']})+e^{\imath t} P_{[\theta,\theta']}]_1
\;,
$$
which is precisely the definition of $\beta$. Now the result follows from the homomorphism property of the Chern numbers and Theorem 5.4.1 in \cite{PS}, or alternatively just a direct calculation. 
\hfill $\Box$

\vspace{.2cm}

Let us point out (as in \cite{RLBL}) that the following situation is possible: the Floquet operator has two gaps and two bands, both of which have vanishing Chern numbers; nevertheless, the two odd invariants $\Ch_{\{0\}\cup I} (V_{\theta'})$ and $\Ch_{\{0\}\cup I} (V_{\theta})$ can be non-trivial according to Proposition~\ref{prop-WindChLink}. This is precisely the situation in the Chalker-Coddington model \cite{DFT}, see Section~\ref{sec-TQW}. In the following section, we shall show that these non-trivial invariants also lead to edge states.

\section{Bulk-boundary correspondence for Floquet systems}
\label{sec-BBC}

The first version of the bulk-boundary correspondence (BBC), Proposition~\ref{prop-ChBBC} below, only invokes the invariants calculated from the orthogonal projections $P_{[\theta,\theta']}$ on the bands of the Floquet operator. This is in complete analogy with the theory of topological insulators \cite{KRS,PS}. The second and new version involving invariants depending on the time evolution is given afterwards in Theorem~\ref{theo-ChBBC3} and Corollary~\ref{coro-ChBBC4}.

\vspace{.2cm}

Let us begin by describing what a Floquet system with boundary is (see \cite{PS}, Section~7.7). The half-space Hamiltonian $\widehat{H}(t)$ is a (Dirichlet) restriction of $H(t)$ to the half-space Hilbert-space $\ell^2(\ZM^{d-1}\times\NM,\CM^L)$. It can be modified by a local boundary condition  depending periodically on time (as in \eqref{eq-HalfSpaceOps}). Associated with it is a time evolution in the Toeplitz extension $T(\Aa_d)$ described in the appendix by
\begin{equation}
\label{eq-TimeEvolEdge}
\imath\,\partial_t \widehat{U}(t)\;=\;\widehat{H}(t)\,\widehat{U}(t)
\;,
\qquad
\widehat{U}(0)\;=\;\widehat{\one}
\;,
\end{equation}
where $\widehat{\one}$ is the identity on the half-space. Again the Floquet operator is $\widehat{U}=\widehat{U}(2\pi)\in T(\Aa_d)$. It is a unitary lift of $U$, namely with the projection $\pi$ onto the bulk as defined in \eqref{eq-ShortExact}, one has $\pi(\widehat{U})=U$. The fact that such a unitary lift indeed exists reflects that $[U]_1=0$. The choice of unitary lift is determined by the boundary conditions included in $\widehat{H}(t)$ and it clearly depends on the whole path $t\in[0,2\pi]\mapsto \widehat{U}(t)$. Actually, one has $\pi(\widehat{U}(t))=U(t)$ for every $t$ due to the locality of $\widehat{H}(t)$. Thus the path $t\in[0,2\pi]\mapsto \widehat{U}(t)$ is a lift of $t\in[0,2\pi]\mapsto {U}(t)$ from the suspension $S\Aa_d$ to the suspension $S\,T(\Aa_d)$ of the Toeplitz algebra. However, if $U=\one$, then $\pi(\widehat{U})=\one$ so that $\widehat{U}\in\Ee_d^+$ lies in the ideal of edge operators, while the path  $t\in[0,2\pi]\mapsto \widehat{U}(t)$ does not. This is a key feature analyzed in detail in the proof of Theorem~\ref{theo-ChBBC3} below.

\vspace{.2cm}

Let us now first suppose that there are two gaps $e^{\imath\theta},e^{\imath\theta'}\not\in\sigma(U)$ with $0\leq \theta<\theta'<2\pi$ with a spectral projection $P_{[\theta,\theta']}\in\Aa_d$ of $U$ as used in Section~\ref{sec-BulkInv}. Due to the existence of the gaps, this spectral projection can be written as a smooth function $G_{\theta,\theta'}:\SM^1\to [0,1]$ of $U$:
$$
P_{[\theta,\theta']}
\;=\;
G_{\theta,\theta'}(U)
\;.
$$
Note that necessarily $(G_{\theta,\theta'})'$ is supported in the two gaps of $U$. This representation leads to a lift $G_{\theta,\theta'}(\widehat{U})$  of $P_{[\theta,\theta']}$ into the Toeplitz extension which can be used to write out the image of the exponential map $\Exp:K_0(\Aa_d)\to K_1(\Ee_d)$, similar as in \cite{KRS,PS}:
\begin{equation}
\label{eq-ExpMap}
\Exp [P_{[\theta,\theta']}]_0\;=\;[e^{-2\pi\imath \,G_{\theta,\theta'}(\widehat{U})}]_1
\;.
\end{equation}
To extract a topological number from the r.h.s., we will need to take a suitable trace $\widetilde{\Tt}$ of half-space operators. It is given by the trace per unit volume on $\ZM^{d-1}$ combined with the usual trace in the $d$-th direction, see Appendix~\ref{app-alg}. Now for a differentiable unitary $\widetilde{V}\in\Ee_d^+$ such that $\widetilde{V}-\one$ is $\widetilde{\Tt}$-traceclass, the boundary invariants associated to an index set $J\subset\{1,\ldots,d-1\}$ of odd cardinality $m$ are defined similarly as in \eqref{OddBulkChernNumbers} by ({\it cf.} Section~5.3 in \cite{PS})
$$
\widetilde{\Ch}_J (\widetilde{V})  
\;=\;   
\frac{\imath\,(\imath \pi)^\frac{m-1}{2}}{m!!}
\sum_{\rho\in \Ss_{J}}(-1)^{\rho}
\;\widetilde{\Tt} \Big ( (\widetilde{V}^* - \one) \nabla_{\rho_1}\widetilde{V}
(\nabla_{\rho_2}\widetilde{V}^* \nabla_{\rho_3}\widetilde{V}) \cdots (\nabla_{\rho_{m-1}}\widetilde{V}^*\nabla_{\rho_m}\widetilde{V}) \Big)
\;.
$$
Note, however, that other than in \eqref{OddBulkChernNumbers} no time derivative is involved here. Now the BBC of \cite{KRS} or Theorem 5.5.1 in \cite{PS} immediately implies:

\begin{proposi}
\label{prop-ChBBC}
Let $J\subset\{1,\ldots,d-1\}$ be an index set of odd cardinality. For $e^{\imath\theta},e^{\imath\theta'}\not\in\sigma(U)$ in two gaps of the quasi-energy spectrum with $0\leq \theta<\theta'<2\pi$, one has
\begin{equation}
\label{eq-ChBBC}
\Ch_{J\cup \{d\}} ( P_{[\theta,\theta']})
\;=\;
\widetilde{\Ch}_J(e^{-2\pi\imath \,G_{\theta,\theta'}(\widehat{U})})
\;.
\end{equation}
\end{proposi}

Let us now give an interpretation of the boundary invariants in dimension $d=2$  by imitating the definition of edge channels for quantum Hall systems \cite{SKR,KRS,EG}.

\begin{defini}
\label{def-EdgeChannel}
Let $\widehat{U}$ be a half-space Floquet operator in physical dimension $d=2$. If $e^{\imath\theta}$ is in a gap of the associated Floquet operator $U$, then the number $N_\theta$ of (oriented) edge bands at $e^{\imath\theta}$ is defined by
\begin{equation}
\label{eq-Edge}
N_\theta
\;=\;
-\,2\pi\imath\;
\widetilde{\Tt}\big((G_\theta)'(\widehat{U})\,\widehat{U}^*\,\nabla_1 \widehat{U}\big)
\;,
\end{equation}
where $(G_\theta)':\SM^1\to[0,\infty)$ is a smooth function supported in the gap of $U$ of unit integral, i.e.
$$
\int_0^{2\pi}
d\varphi
\;(G_\theta)'(e^{\imath\varphi})
\;=\;1
\;.
$$
\end{defini}

Let us note that, first of all, the operator $(G_\theta)'(\widehat{U})$ is indeed $\widetilde{\Tt}$-traceclass \cite{PS} so that the r.h.s. of \eqref{eq-Edge} makes sense. Second of all, the r.h.s. is indeed independent of the particular choice of $(G_\theta)'$ and it is an integer number. This follows from the proof of Proposition~\ref{prop-ChBBC2} below.

\begin{proposi}
\label{prop-ChBBC2}
Let $d=2$. For $e^{\imath\theta},e^{\imath\theta'}\not\in\sigma(U)$ in two gaps of the quasi-energy spectrum of $U$, one has
\begin{equation}
\label{eq-ChBBC2}
\Ch_{\{1,2\}} ( P_{[\theta,\theta']})
\;=\;
N_{\theta}\,-\,N_{\theta'}
\;.
\end{equation}
\end{proposi}

\noindent {\bf Proof.} (This follows closely Proposition 7.1.2 in \cite{PS}) Let us start from the r.h.s. of \eqref{eq-ChBBC}. As $d=2$, one necessarily has $J=\{1\}$ so that for any integer $n\not=0$ by the additivity of the odd Chern number,
$$
\Ch_{\{1,2\}} ( P_{[\theta,\theta']})
\;=\;
\widetilde{\Ch}_{\{1\}} (e^{-2\pi\imath \,G_{\theta,\theta'}(\widehat{U})})
\;=\;
\frac{1}{n}\,
\widetilde{\Ch}_{\{1\}} (e^{-2\pi\imath n\,G_{\theta,\theta'}(\widehat{U})})
\;.
$$
Writing this out explicitly, one obtains using the cyclicity of the trace
\begin{align*}
\Ch_{\{1,2\}} ( P_{[\theta,\theta']})
& \;=\;
\frac{\imath}{n}
\;
\widetilde{\Tt}
\big(
(e^{2\pi\imath n\,G_{\theta,\theta'}(\widehat{U})}-\one)\,\nabla_1 e^{-2\pi\imath n\,G_{\theta,\theta'}(\widehat{U})}
\big)
\\
& \;=\;
2\pi
\;
\widetilde{\Tt}
\big(
(\one-e^{-2\pi\imath n\,G_{\theta,\theta'}(\widehat{U})})\,\nabla_1 G_{\theta,\theta'}(\widehat{U})
\big)
\\
& \;=\;
-\,2\pi\imath
\;
\widetilde{\Tt}
\big(
(\one-e^{-2\pi\imath n\,G_{\theta,\theta'}(\widehat{U})}) 
\, (G_{\theta,\theta'})'(\widehat{U})\,\widehat{U}^*\,\nabla_1 \widehat{U}
\big)
\;,
\end{align*}
where
$$
(G_{\theta,\theta'})'(e^{\imath\varphi})
\;=\;
\partial_\varphi\,\big(G_{\theta,\theta'}(e^{\imath\varphi})\big)
\;=\;\imath\,u\,\partial_u\,\big(G_{\theta,\theta'}(u) \big)
$$
As this holds for all $n\not=0$, a Fourier series argument as at the end of the proof of Proposition~7.1.2 in \cite{PS} shows
$$
\Ch_{\{1,2\}} ( P_{[\theta,\theta']})
\;=\;
-\,2\pi\imath
\;
\widetilde{\Tt}
\big(
 (G_{\theta,\theta'})'(\widehat{U})\,\widehat{U}^*\,\nabla_1 \widehat{U}
\big)
\;.
$$
Finally recall that $(G_{\theta,\theta'})'$ is supported by two gaps of $U$ and has integrals equal to $1$ and $-1$ on these gaps. Thus, it can  be written as
$$
(G_{\theta,\theta'})'
\;=\;
(G_{\theta})'\,-\,(G_{\theta'})'
\;,
$$
with two functions $(G_{\theta})'$ and $(G_{\theta'})'$ as appearing in Definition~\ref{def-EdgeChannel}. This concludes the proof.
\hfill $\Box$

\vspace{.2cm}

Let us now analyze the odd Chern numbers $\Ch_{\{0\}\cup I} (V_\theta)$ defined in \eqref{OddBulkChernNumbers} and appearing in Proposition~\ref{prop-WindChLink}. For that purpose, we start by considering the path $ t\in[0,2\pi]\mapsto V_\theta(t)$ specifying a class $[V_\theta]_1\in K_1(S\Aa_d)$. One way to proceed is to apply the inverse of the Bott map $\beta^{-1}:K_1(S\Aa_d)\to K_0(\Aa_d)$  to obtain a projection class $[P]_0=\beta^{-1}([V_\theta]_1)\in K_0(\Aa_d)$ and then to apply the exponential map $\Exp:K_0(\Aa_d)\to K_1(\Ee_d)$ similar as in \eqref{eq-ExpMap} to produce a possibly non-trivial unitary in the edge algebra. The problem with approach this is that the map $\beta^{-1}$ is quite involved (see the Atiyah-Bott proof of the surjectivity of $\beta$). On the other hand, one can make use of the commutative diagram 
\begin{diagram}
& K_1(S \Aa_d)  &\rTo{ \ \ \beta^{-1} \ \ } \ \  & K_0(\Aa_d) \\
&\dTo{{\rm Ind} }      &                                            &\dTo{{\rm Exp} }                  \\
&K_0(S\Ee_{d})   &\rTo{\ \ \Theta^{-1} \ \ }  & K_1( \Ee_{d}) 
\end{diagram}
to calculate $\Exp\circ\beta^{-1}=\Theta^{-1}\circ\Ind$. As we shall see, the evaluation of the index map is straightforward, and also the calculation of the inverse suspension map $\Theta^{-1}$ is possible, {\it e.g.} as in Theorem~4.1.9 in \cite{PS}. 

\begin{theo}
\label{theo-ChBBC3} 
Let $e^{\imath\theta}\not\in\sigma(U)$ be in a gap of the Floquet operator and let $\theta'$ be a point not lying in that gap. Further let $G_\theta:\SM^1\to[0,1]$ be a smooth function except at $e^{\imath\theta'}$ such that $G'_\theta$ is as in Definition~\ref{def-EdgeChannel}. Then
$$
\Theta^{-1}(\Ind ([V_\theta]_1))
\;=\;
[e^{-2\pi\imath \,G_{\theta}(\widehat{U})}]_1
\;.
$$
\end{theo}

\noindent {\bf Proof.} Let us note that $G_{\theta}(\widehat{U})$ is not in $T(\Aa_d)$ because $G_{\theta}$ is discontinuous and has a jump of $1$ at $e^{\imath\theta'}$. However, $e^{-2\pi\imath \,G_{\theta}(\widehat{U})}$ is a continuous function of $\widehat{U}$ and thus in $T(\Aa_d)$. Moreover, the smooth function $f_\theta(e^{\imath\varphi})=e^{-2\pi\imath \,G_{\theta}(e^{\imath\varphi})}-1$ is entirely supported by the gap of $U$. As $\widehat{U}$ is a lift of $U$, this implies that $e^{-2\pi\imath \,G_{\theta}(\widehat{U})}-\widehat{\one}$ is indeed a $\widetilde{\Tt}$-traceclass element of $\Ee_d$ \cite{EG,PS}. Note that here $\widehat{\one}$ is the adjoint unit in $\Ee_d^+$, which is identified with identity operator on the upper half-space. Thus indeed $[e^{-2\pi\imath \,G_{\theta}(\widehat{U})}]_1\in K_1(\Ee_d)$. 

\vspace{.1cm}

Furthermore, the spectrum of the operators $Ue^{\imath s h_\theta}=e^{\imath (s-2\pi) h_\theta}$ lies outside of the support of $f_\theta$ for all $s\in[0,2\pi]$. Hence its lift $\widehat{U}e^{\imath s \hat{h}_\theta}$, which is {\em not} equal to $e^{\imath (s-2\pi) \hat{h}_\theta}$, also leads to $\widetilde{\Tt}$-traceclass elements $e^{-2\pi\imath \,G_{\theta}(\widehat{U}e^{\imath s \hat{h}_\theta})}$. This can be used as a homotopy to establish
\begin{equation}
\label{eq-ExpEq}
[e^{-2\pi\imath \,G_{\theta}(\widehat{U})}]_1
\;=\;
[e^{-2\pi\imath \,G_{\theta}(\widehat{U}e^{2\pi\imath  \hat{h}_\theta})}]_1
\;.
\end{equation}
But the solution to
$$
\imath\,\partial_t \widehat{U}_\theta(t)\;=\;\widehat{H}_\theta(t)\,\widehat{U}_\theta(t)
\;,
\qquad
\widehat{U}_\theta(0)\;=\;\widehat{\one}
\;,
$$
where $\widehat{H}_\theta(t)$ is the half-space restriction of ${H}_\theta(t)$ given by \eqref{eq-HamTheta}, is given by
$$
\widehat{U}_\theta(t)
\;=\;
\left\{
\begin{array}{cc}
\widehat{U}(2t)\;,
& 
t\in[0,\pi]\;,
\\
\widehat{U}\,e^{2(t-\pi)\imath \,\hat{h}_\theta}\;,
&
t\in(\pi,2\pi]
\;.
\end{array}
\right.
$$
In particular, $\widehat{U}_\theta(2\pi)=\widehat{U}e^{2\pi\imath  \hat{h}_\theta}$. Replacing this in \eqref{eq-ExpEq} and using the time evolution \eqref {eq-PeriTimeEvol} instead of \eqref{eq-TimeEvol} shows that it is sufficient to consider the special case $U=U(2\pi)=\one$. Hence from now on $V(t)=U(t)$ and there are no further indices $\theta$. Let us stress that in the following $\widehat{V}(t)=\Pi V(t) \Pi^*$ is the contraction lift, but $\widehat{U}(t)$ is the unitary solution of \eqref{eq-TimeEvolEdge}, and these objects are {\em not} the same.

\vspace{.1cm}

As $U=U(2\pi)=\one$, also $\widehat{U}-\widehat{\one}\in\Ee_d$ and there exists a sequence of smooth functions as above converging to the continuous function $G$ with jump $1$ at $1=e^{\imath 0}$, given by $G(e^{\imath \varphi})=\frac{\varphi}{2\pi}$. For this function, 
$$
e^{-2\pi\imath \,G(\widehat{U})}-\widehat{\one}
\;=\;
\widehat{U}^*-\widehat{\one}
\;.
$$
Hence one has to show $\Theta^{-1}(\Ind ([V]_1))=[\widehat{U}^*]_1$.

\vspace{.1cm}

For the calculation of the image of the index map $\Ind ([V]_1)\in K_0(S\Ee_d)$ one usually uses the fact that the diagonal operator $\diag({V}(t),{V}(t)^*)$ has a unitary lift to $M_2(T(\Aa_d))$, the $2\times 2$ matrices with entries from $T(\Aa_d)$. This lift can be used, and will be used in the proof of Theorem~\ref{theo-ThetaInd} below. However, there is another lift which from a physical perspective is more natural, namely one can simply use the unitary ${V}(t)$ on the full space and decompose it into upper and lower half:
$$
V(t)
\;=\;
\begin{pmatrix}
\widehat{V}(t) & \Pi V(t)(\Pi^c)^* \\ \Pi^c V(t)\Pi^* & \Pi^c V(t)(\Pi^c)^*
\end{pmatrix}
\;,
$$
where $\Pi^c:\ell^2(\ZM^d,\CM^L)\to \ell^2(\ZM^{d-1}\times \NM_-,\CM^L)$ is the partial isometry onto the lower half (here $\NM=\{0,1,2,\ldots\}$ and $\NM_-=\{-1,-2,\ldots\}$). One problem is that three of the matrix entries of $V(t)$ are not in $T(\Aa_d)$ (or more precisely, are not representatives of elements in $T(\Aa_d)$). It is, however, possible to reflect the lower half to the upper half via  $R:\ell^2(\ZM^{d-1}\times \NM,\CM^L)\to \ell^2(\ZM^{d-1}\times \NM_-,\CM^L)$ defined by
\begin{equation}
\label{eq-Reflect}
R|n_1,n_2\rangle
\;=\;
|-n_1,-n_2-1\rangle
\;,
\qquad
n_1\in\ZM^{d-1}\;,\;\;
n_2\in\NM
\;.
\end{equation}
This reflection satisfies $R^*R=\widehat{\one}=\Pi\Pi^*$ and $RR^*=\Pi^c(\Pi^c)^*$. Then for $A\in\Aa_d$, one has $R^*\Pi^c A(\Pi^c)^*R\in T(\Aa_d)$, however, in a covariant representation with partially reversed magnetic translations. Some further details on this and $R$ is given in Appendix~\ref{app-alg}. Consequently, the following is a unitary operator in $M_2(T(\Aa_d))$ (more precisely, a representative of such an operator):
\begin{equation}
\label{eq-What}
\widehat{W}(t)
\;=\;
\begin{pmatrix} 
\widehat{\one} & 0 \\ 0 & R^*
\end{pmatrix}
V(t)
\begin{pmatrix} 
\widehat{\one} & 0 \\ 0 & R
\end{pmatrix}
\;=\;
\begin{pmatrix}
\widehat{V}(t) & \Pi V(t)(\Pi^c)^*R \\ R^*\Pi^c V(t)\Pi^* & R^*\Pi^c V(t)(\Pi^c)^*R
\end{pmatrix}
\;.
\end{equation}
It satisfies $\widehat{W}(0)=\widehat{W}(2\pi)=\one=\diag(\widehat{\one},\widehat{\one})$, as well as the differential equation
$$
\imath\;\partial_t\widehat{W}(t)
\;=\;
\imath\;\partial_t\,
\begin{pmatrix} 
\widehat{\one} & 0 \\ 0 & R^*
\end{pmatrix}
V(t)
\begin{pmatrix} 
\widehat{\one} & 0 \\ 0 & R
\end{pmatrix}
\;=\;
\begin{pmatrix} 
\widehat{\one} & 0 \\ 0 & R^*
\end{pmatrix}
\imath\;\partial_t\,
V(t)
\begin{pmatrix} 
\widehat{\one} & 0 \\ 0 & R
\end{pmatrix}
\;=\;
\breve{H}(t)\,\widehat{W}(t)
\;,
$$
where we set
\begin{equation}
\label{eq-HamRefl}
\breve{H}(t)\;=\;
\begin{pmatrix} 
\widehat{\one} & 0 \\ 0 & R^*
\end{pmatrix}
H(t)
\begin{pmatrix} 
\widehat{\one} & 0 \\ 0 & R
\end{pmatrix}
\;\in\;M_2(T(\Aa_d))
\;.
\end{equation}
Moreover, $\widehat{W}(t)$ is indeed a suitable lift for the calculation of the index map because it is unitary and projects down to a diagonal unitary with the upper left entry equal to $V(t)$. (Note that it is not necessary that the lower right entry is $V^*$, see \cite{WO}, but in the proof of Theorem~\ref{theo-ThetaInd} below this standard choice is used.) Thus \cite{RLL,WO,PS}
$$
\Ind ([V]_1)
\;=\;
\left[
\widehat{W}(t)
\begin{pmatrix}
\widehat{\one} & 0 \\ 0 & 0 
\end{pmatrix}
\widehat{W}(t)^*
\right]_0
-
\left[
\begin{pmatrix}
\widehat{\one} & 0 \\ 0 & 0 
\end{pmatrix}
\right]_0
\;.
$$
The projection in the first bracket can be evaluated
$$
\widetilde{P}(t)
\;=\;
\widehat{W}(t)
\begin{pmatrix}
\widehat{\one} & 0 \\ 0 & 0 
\end{pmatrix}
\widehat{W}(t)^*
\;=\;
\begin{pmatrix}
\widehat{V}(t)\widehat{V}^*(t) & \widehat{V}(t) \Pi V(t)^*(\Pi^c)^*R \\ R^*\Pi^c V(t)\Pi^*\widehat{V}^*(t) & R^*\Pi^c V(t)\Pi\Pi^*V(t)^*(\Pi^c)^*R
\end{pmatrix}
\;.
$$
Even though $\widehat{W}(t)$ is only in $M_2(T(\Aa_d))$, this formula shows that $\widetilde{P}(t)$ lies in $M_2(\Ee_d^+)$ because $R^*\Pi^c V(t)\Pi^*$ lies in $\Ee_d$.

\vspace{.1cm}

Given the loop $t\in[0,2\pi]\mapsto\widetilde{P}(t)\in M_2(\Ee_d^+)$, let us now calculate the associated adiabatic evolution as in Section~4.1.5 in \cite{PS}:
$$
\imath\,\partial_t \widetilde{W}(t)
\;=\;
\big(\widetilde{A}(t)\,+\,\imath [\partial_t \widetilde{P}(t),\widetilde{P}(t)]\big)\widetilde{W}(t)
\;,
\qquad
\widetilde{W}(0)\,=\,
\begin{pmatrix}
\widehat{\one} & 0 \\ 0 & \widehat{\one} 
\end{pmatrix}
\;,
$$
where $\widetilde{A}(t)=\widetilde{A}(t)^*\in M_2(\Ee_d^+)$ satisfying $[\widetilde{A}(t),\widetilde{P}(t)]=0$ can be chosen freely, but this freedom will not be used later on, that is, we simply set $\widetilde{A}(t)=0$.  The solution $\widetilde{W}(t)$ is a unitary in $M_2(\Ee_d^+)$ which according to Proposition~4.1.8 in \cite{PS} -- nothing but the standard result on the adiabatic evolution of a path of projections -- is given by
$$
\widetilde{P}(t)
\;=\;
\widetilde{W}(t)
\widetilde{P}(0)
\widetilde{W}(t)^*
\;=\;
\widetilde{W}(t)
\begin{pmatrix}
\widehat{\one} & 0 \\ 0 & 0 
\end{pmatrix}
\widetilde{W}(t)^*
\;.
$$
Some word of caution: one should {\em not} conclude that $\widetilde{W}(t)$ is equal to $\widehat{W}(t)$. Indeed, $\widetilde{W}(t)\in M_2(\Ee_d^+)$ lies in the unitized edge algebra, but $\widehat{W}(t)$ is only a half-space operator in $M_2(T(\Aa_d))$. Now as $\widetilde{P}(0)=\widetilde{P}(2\pi)$, it follows that $\widetilde{W}(2\pi)=\diag(u,v)$ is diagonal with two unitaries $u$ and $v$ in $\Ee_d^+$. By Theorem~4.1.9 in \cite{PS} one now has
\begin{equation}
\label{eq-IndRes}
\Theta^{-1}(\Ind ([V]_1))
\;=\;[u]_1
\;.
\end{equation}
Hence we need to calculate the solution of the adiabatic equation and show that $u$ is homotopic to $\widehat{U}^*$ in $\Ee_d^+$. 

\vspace{.1cm}

For the homotopy argument, let us modify the Hamiltonian in \eqref{eq-HamRefl} as follows:
\begin{equation}
\label{eq-HamDecop}
\breve{H}_\lambda(t)
\;=\;
\begin{pmatrix} 
\widehat{H}(t) & \lambda\, k(t) \\ \lambda \,k(t)^* & \widehat{H}'(t)
\end{pmatrix}
\;=\;
\begin{pmatrix} 
\widehat{H}(t) & 0 \\ 0 & \widehat{H}'(t)
\end{pmatrix}
\;+\;\lambda\,
K(t)
\;,
\end{equation}
where $k(t)\in\Ee_d$ and $K(t)=K(t)^*\in M_2(\Ee_d)$ couple lower and upper halves, and $\widehat{H}'(t)\in T(\Aa_d)$ is the (reflected) Hamiltonian on the lower half. Then $\breve{H}_1(t)=\breve{H}(t)$, and $\breve{H}_0(t)$ is diagonal. Associated with $\breve{H}_\lambda(t)$, there is the unitary solution 
\begin{equation}
\label{eq-WHamCechLam}
\imath\;\partial_t\widehat{W}_\lambda(t)
\;=\;
\breve{H}_\lambda(t)\,\widehat{W}_\lambda(t)
\;,
\qquad
\widehat{W}_\lambda(0)
\;=\;
\one\;=\;
\begin{pmatrix}
\widehat{\one} & 0 \\ 0 & \widehat{\one} 
\end{pmatrix}
\;.
\end{equation}
Due to the change in the dynamics, $t\in[0,2\pi]\mapsto\widehat{W}_\lambda(t)$ will {\em not} be a loop any longer. Crucial is, however, that the perturbation $K(t)$ is in $\Ee_d$ and therefore by a standard perturbative argument (DuHamel's formula) also the end point $\widehat{W}_\lambda(2\pi)$ differs from $\widehat{W}_1(2\pi)=\one$ merely by an operator from $\Ee_d$. Thus $\widehat{W}_\lambda(2\pi)\in M_2(\Ee_d^+)$ for all $\lambda$. Also note that
$$
\widehat{W}_0(2\pi)
\;=\;
\begin{pmatrix}
\widehat{U} & 0 \\ 0 & \widehat{U}'
\end{pmatrix}
\;,
$$
for some unitary $\widehat{U}'\in\Ee_d^+$ which will not be used below, but is essentially the Floquet operator of the lower half. From $\widehat{W}_\lambda(t)$ one gets
$$
\widetilde{P}_\lambda(t)
\;=\;
\widehat{W}_\lambda(t)
\begin{pmatrix}
\widehat{\one} & 0 \\ 0 & 0 
\end{pmatrix}
\widehat{W}_\lambda(t)^*
\;,
$$
which by locality of $\breve{H}_\lambda(t)$ is a path in $M_2(\Ee_d^+)$, but this is also not a loop any longer. Let us consider the associated adiabatic equation
\begin{equation}
\label{eq-AdEq}
\imath\,\partial_t \widetilde{W}_\lambda(t)
\;=\;
\imath [\partial_t \widetilde{P}_\lambda(t),\widetilde{P}_\lambda(t)]\,\widetilde{W}_\lambda(t)
\;,
\qquad
\widetilde{W}_\lambda(0)\,=\,
\begin{pmatrix}
\widehat{\one} & 0 \\ 0 & \widehat{\one} 
\end{pmatrix}
\;.
\end{equation}
For the calculation of the commutator, let us note that
$$
\imath\;\partial_t\widetilde{P}_\lambda(t)
\;=\;
\imath\;\partial_t\,\big(\widehat{W}_\lambda(t)\,\widetilde{P}_\lambda(0)\,\widehat{W}_\lambda(t)^*\big)
\;=\;
\breve{H}_\lambda(t)\widetilde{P}_\lambda(t)-\widetilde{P}_\lambda(t)\breve{H}_\lambda(t)
\;,
$$
and thus, still with $\one=\diag(\widehat{\one},\widehat{\one})$,
\begin{align*}
\imath\,[\partial_t \widetilde{P}_\lambda(t),\widetilde{P}_\lambda(t)]
& \;=\;
\imath\,\partial_t \widetilde{P}_\lambda(t) \,\widetilde{P}_\lambda(t)
\;-\;
\imath\,\widetilde{P}_\lambda(t)\,\partial_t \widetilde{P}_\lambda(t)
\\
& \;=\;
(\one -\widetilde{P}_\lambda(t))\,\imath\,\partial_t \widetilde{P}_\lambda(t) \,\widetilde{P}_\lambda(t)
\;-\;
\widetilde{P}_\lambda(t)\,\imath\,\partial_t \widetilde{P}_\lambda(t) \,(\one -\widetilde{P}_\lambda(t))
\\
& 
\;=\;
(\one-\widetilde{P}_\lambda(t))\,\breve{H}_\lambda(t)\,\widetilde{P}_\lambda(t)
\;+\;
\widetilde{P}_\lambda(t)\,\breve{H}_\lambda(t)\,(\one-\widetilde{P}_\lambda(t))
\;.
\end{align*}
In particular, this shows that the commutator is in $\Ee_d$ and therefore also the solution $\widetilde{W}_\lambda(t)$ of \eqref{eq-AdEq} lies in $M_2(\Ee_d^+)$ for all $t$. For $\lambda=0$, one has $\widetilde{P}_0(t)=\diag(\widehat{\one},0)$ and therefore $\widetilde{W}_0(t)=\one$ for all $t$.

\vspace{.2cm}

We are interested in the upper left entry of $\widetilde{W}_1(2\pi)=\widetilde{W}(2\pi)=\diag(u,v)$. To connect it to $\widehat{U}^*$, let us consider the evolution of $\widehat{W}_\lambda(t)^* \widetilde{W}_\lambda(t)$. From \eqref{eq-WHamCechLam} and \eqref{eq-AdEq}, one deduces
$$
\imath\;\partial_t\, \big(\widehat{W}_\lambda(t)^* \widetilde{W}_\lambda(t)\big)
\;=\;
-\,\widehat{W}_\lambda(t)^* \big(\breve{H}_\lambda(t)-\widetilde{H}_\lambda(t)\big)\widehat{W}_\lambda(t)\; 
\widehat{W}_\lambda(t)^* \widetilde{W}_\lambda(t)
\;,
$$
as well as $\widehat{W}_\lambda(0)^* \widetilde{W}_\lambda(0)=\one$. However,
$$
\breve{H}_\lambda(t)-\widetilde{H}_\lambda(t)
\;=\;
\widetilde{P}_\lambda(t) \breve{H}_\lambda(t) \widetilde{P}_\lambda(t)\,+\,(\one-\widetilde{P}_\lambda(t)) \,\breve{H}_\lambda(t)\, 
(\one-\widetilde{P}_\lambda(t))
\;,
$$
and therefore
\begin{align*}
 \widehat{W}_\lambda(t)^* \,&(\breve{H}_\lambda(t)-\widetilde{H}_\lambda(t))\,\widehat{W}_\lambda(t)
\\
& \;=\;
\widetilde{P}_\lambda(0) \,\widehat{W}_\lambda(t)^* \,\breve{H}_\lambda(t) \,\widehat{W}_\lambda(t) \,\widetilde{P}_\lambda(0)\,+\,
(\one-\widetilde{P}_\lambda(0)) \,\widehat{W}_\lambda(t)^* \,\breve{H}_\lambda(t)\, \widehat{W}_\lambda(t) \,(\one-\widetilde{P}_\lambda(0))
\end{align*}
is diagonal. Hence $\widehat{W}_\lambda(t)^* \widetilde{W}_\lambda(t)$ is diagonal for all $t\in[0,2\pi]$ and $\lambda\in[0,1]$. Furthermore, $\widehat{W}_\lambda(2\pi)^* \widetilde{W}_\lambda(2\pi)$ is in $M_2(\Ee_d^+)$ for all $\lambda$ because each factor is. Thus this provides a homotopy between $\widehat{W}_1(2\pi)^* \widetilde{W}_1(2\pi)=\widetilde{W}(2\pi)=\diag(u,v)$ and $\widehat{W}_0(2\pi)^* \widetilde{W}_0(2\pi)=\diag(\widehat{U},\widehat{U}')^*
$ inside the diagonal operators in $M_2(\Ee_d^+)$. In particular, $u$ is homotopic to $\widehat{U}^*$ in the unitaries in $\Ee_d^+$. This finishes the proof of \eqref{eq-IndRes} in the special case $U(2\pi)=\one$ which by the above arguments implies the general case.
\hfill $\Box$

\vspace{.2cm}

\noindent {\bf Proof of Theorem~\ref{theo-ThetaInd}.} As in the above proof, let $V(t)=U(t)\in\Aa$. Let then $\widehat{V}(t)\in\Bb$ be a contraction lift of $V(t)$. A unitary lift of $W(t)=\diag(V(t),V(t)^*)$ is now the Halmos dilation of $\widehat{V}(t)$:
$$
\widehat{W}(t)
\;=\;
\begin{pmatrix}
\widehat{V}(t) & (\widehat{\one} -\widehat{V}(t)\widehat{V}(t)^*)^{\frac{1}{2}} 
\\ 
(\widehat{\one} -\widehat{V}(t)^*\widehat{V}(t))^{\frac{1}{2}} & \widehat{V}(t)^*
\end{pmatrix}
\;.
$$
This replaces \eqref{eq-What}. Because $\imath\,\partial_t{W}(t)=\diag({H}(t),-{H}(t))\,{W}(t)$, one has again a differential equation $\imath\,\partial_t\widehat{W}(t)= \breve{H}(t)\,\widehat{W}(t)$ where now $\breve{H}(t)=\diag(\widehat{H}(t),-\widehat{H}(t))+K(t)$ with another $K(t)\in M_2(\Ee)$. The coupling term $K(t)$ can be tuned down as in \eqref{eq-HamDecop} and this then leads to $\widehat{W}_\lambda(t)$ as in \eqref{eq-WHamCechLam}. Now all elements are in place to repeat  the proof above.
\hfill $\Box$

\vspace{.2cm}

Theorem~\ref{theo-ChBBC3} directly implies links of the bulk invariant to a boundary invariants. For $d=2$ this proves the claims of \cite{RLBL}.

\begin{coro}
\label{coro-ChBBC4} 
Under the same assumptions as in Theorem~\ref{theo-ChBBC3}, one has for any index set $J\subset\{1,\ldots,d-1\}$ of odd cardinality
$$
\Ch_{\{0\}\cup J\cup\{d\}} (V_\theta)
\;=\;
\widetilde{\Ch}_{J}\big(e^{-2\pi\imath \,G_{\theta}(\widehat{U})}\big)
\;.
$$
For dimension $d=2$, 
$$
\Ch_{\{0,1,2\}} (V_\theta)
\;=\;
N_\theta
\;.
$$
\end{coro}

\noindent {\bf Proof.}
The duality results Theorems 5.5.1 and 5.4.1 in \cite{PS} show
$$
\Ch_{\{0\}\cup J\cup\{d\}} (V_\theta)
\;=\;
\widetilde{\Ch}_{\{0\}\cup J} \big(\Ind ([V_\theta]_1)\big)
\;=\;
\widetilde{\Ch}_{J} \big(\Theta^{-1}(\Ind ([V_\theta]_1))\big)
\;,
$$
so that replacing Theorem~\ref{theo-ChBBC3} shows the first claim. The second claim follows by the same argument as in the proof of Proposition~\ref{prop-ChBBC2}. 
\hfill $\Box$

\section{Topological quantum walks}
\label{sec-TQW}

A $d$-dimensional quantum walk is given by a succession of unitary operators $U(1),\ldots,U(N)$ on $\ell^2(\ZM^d,\CM^L)$, which are obtained by what is often called a protocol \cite{Kit}. Their product then defines the analog of the Floquet operator of a periodically driven system:
\begin{equation}
\label{eq-QWFloquet}
U\;=\;U(N)\cdots U(1)
\;.
\end{equation}
Similarly as in \cite{DFT}, the following assumptions are imposed here:
\begin{itemize}

\item[{\rm (i)}]  For $n=1,\ldots,N$, $U(n)\in\Aa_d$ is differentiable w.r.t. $\nabla_1,\ldots,\nabla_d$.

\item[{\rm (ii)}] $U$ is insulating in the sense that it has gaps in its spectrum.

\item[{\rm (iii)}] For each $n=1,\ldots,N$ there is given $H(n)\in\Aa_d$  such that $U(n)=e^{2\pi\imath H(n)}$. 

\end{itemize}
The hypothesis (i) means that each $U(n)$ is local (on the $d$-dimensional lattice), and (iii) states that each individual step $U(n)$ is topologically trivial as it specifies a path $s\in[0,1]\mapsto e^{2\pi\imath s H(n)}$ of unitaries connecting $U(n)$ to the identity. If $U(n)$ is gapped, an effective Hamiltonian $H(n)$ is given by taking a suitable logarithm of $U(n)$. However, let us stress that there are several choices possible and that they may alter the topology of the quantum walk. Hence the hypothesis is that one such choice has been made and is natural, see the discussions in \cite{DFT,NR}. Let us note that, even if $U(n)$ is a finite range operator, the effective Hamiltonian $H(n)$ typically is not of finite range. There may, however, be other paths $s\in[0,1]\mapsto U(n)_s$ of unitaries from $\one$ to $U(n)$ which are homotopic to the above path in $\Aa_d$ and which are of finite range. 

\vspace{.2cm}

From the above data, one can now define a time-dependent $2\pi$-periodic Hamiltonian:
\begin{equation}
\label{eq-QWHam}
H(t)
\;=\;
\sum_{n=1}^N
\;
N\;H(n)\;\chi\big(t\in [2\pi\tfrac{n-1}{N},2\pi\tfrac{n}{N})\big)
\;.
\qquad
t\in[0,2\pi)
\;,
\end{equation}
where $\chi$ denotes the indicator function. By construction the associated Floquet operator coincides with $U$ given by \eqref{eq-QWFloquet}. Consequently, the definitions and facts from periodically driven systems as developed in Section~\ref{sec-BulkInv}  can be directly applied. Also the bulk-boundary correspondence given in Section~\ref{sec-BBC} holds, but the choice of the half-space quantum walk deserves some discussion. Of course, one can work with the half-space reduction $\widehat{H}(t)$ of \eqref{eq-QWHam}, and then \eqref{eq-TimeEvolEdge} does provide $\widehat{U}(t)$ and $\widehat{U}=\widehat{U}(2\pi)$, but even if $U$ was finite range, this typically does {\it not} lead to a finite range half-space quantum walk $\widehat{U}$. On the other hand, there are natural ad hoc ways to modify the quantum walk at the boundaries \cite{HC,Kit} and this just reflects that different lifts into the half-space algebra can be chosen. In a given situation, one has to analyze whether this natural construction is homotopic to the above construction.

\vspace{.2cm}

Let us illustrate all the above by dealing with a concrete example, namely the Chalker-Coddington model. We follow closely \cite{HC,DFT}. In the above framework, the Hilbert space is $\ell^2(\ZM^2,\CM^2)$ and there are $4$ steps, two which are invariant under translation and specified by a parameter $\beta\in[0,\pi]$
$$
U(1)
\;=\;
\begin{pmatrix}
\sin(\beta) S_1 & \cos(\beta) \\ \cos(\beta) & -\sin(\beta) S^*_1
\end{pmatrix}
\;,
\qquad
U(3)
\;=\;
\begin{pmatrix}
\cos(\beta) & \sin(\beta) S_2 \\  \sin(\beta) S^*_2 & -\cos(\beta)
\end{pmatrix}
\;,
$$
and two which are diagonal and given by real random processes of i.i.d. angles $\phi_j=(\phi_j(n))_{n\in\ZM^2}$ with $j=1,2,3,4$ and a coupling constant $\lambda\geq 0$:
$$
U(2)
\;=\;
\begin{pmatrix}
e^{\imath \lambda \phi_1} & 0 \\ 0 & e^{\imath \lambda \phi_2} 
\end{pmatrix}
\;,
\qquad
U(4)
\;=\;
\begin{pmatrix}
e^{\imath \lambda \phi_3} & 0 \\ 0 & e^{\imath\lambda  \phi_4} 
\end{pmatrix}
\;.
$$
For $\lambda=0$ and $\beta\not=\frac{\pi}{4}$, the unitary $U$ is gapped and has two bands. This remains true for sufficiently small $\lambda$. As shown in \cite{DFT}, both bands have trivial two-dimensional Chern numbers and hence Proposition~\ref{prop-ChBBC2} does not allow to conclude that there are non-trivial edge channels $N_\theta$ and $N_{\theta'}$ in the two gaps containing $e^{\imath\theta}$ and $e^{\imath\theta'}$. However, it is also shown in \cite{DFT} that $\Ch_{\{0,1,2\}} (V_\theta)=1$ and $\Ch_{\{0,1,2\}} (V_{\theta'})=1$ so that  $N_\theta=N_{\theta'}=1$ by Corollary~\ref{coro-ChBBC4}. These edge bands were first found in \cite{HC}. The results of \cite{DFT} combined with the present paper explain the topological origin of these edge channels and also show their stability against disordered perturbations, in analogy to the situation in the quantum Hall effect \cite{SKR,KRS,EG}.

\appendix

\section{Resum\'e on covariant operator families}
\label{app-alg}

This appendix reviews the C$^*$-algebraic description of covariant operators as introduced by Bellissard \cite{Bel}.   Let us consider operators on a $d$-dimensional lattice Hilbert space $\ell^2(\ZM^d)\otimes\CM^L$ with $L$-dimensional fiber. An orthonormal basis is given by $|n,l\rangle$ with $n=(n_1,\ldots,n_d)\in\ZM^d$ and $l=1,\ldots,L$. The shift operators and position operator are defined by
$$
S_j|n,l\rangle\;=\;|n+e_j,l\rangle
\;,
\qquad
X_j|n,l\rangle\;=\;n_j\,|n,l\rangle
\;,
$$
where $j=1,\ldots,d$ and $e_j$ is the $j$th standard basis vector. If a constant magnetic field is given by a real antisymmetric $d\times d$ matrix $\BB=\BB_+-\BB_-$ with lower triagunlar part $\BB_+$, then the magnetic translations (in the Landau gauge) are (see \cite{PS})
$$
V_j\;=\;
e^{\imath\,\langle X|\BB_+|e_j\rangle}\,S_j
\;,
$$
where $e_j$ is the $j$th vector of the standard basis of $\RM^d$ and the euclidean scalar product has been used. Let now $(\Omega,\tau,\ZM^d,\PM)$ be a compact space equipped with a $\ZM^d$-action $\tau=(\tau_1,\ldots,\tau_d)$ and an invariant and ergodic probability measure $\PM$. It is called the space  of disorder or crystalline configurations.  A finite-range covariant operator family  $A=(A_{\omega})_{\omega\in\Omega}$ consists of bounded operators on $\ell^2(\ZM^d)\otimes\CM^L$, strongly continuous in $\omega$, with $\langle n|A_\omega|n'\rangle=0$ for $|n-n'|>\rho$ for some finite $\rho$, and satisfying the covariance relation
$$
V_j
A_{\omega}
V_j^*
\;=\;
A_{\tau_j \omega}
\;,
\qquad a\in\ZM^d
\;.
$$
Products and adjoints of such families are again covariant operator families. A C$^*$-norm on the short-range covariant families is given by $\| A\| = \sup_{\omega \in \Omega}\| A_{\omega}\|$ and the corresponding closure is the reduced twisted C$^*$-crossed product algebra $\Aa_d=C(\Omega)\rtimes_\BB\ZM^d$. An abstract way to define this algebra $\Aa_d$ is to say that it is generated by $d$ unitaries $u_1,\ldots,u_d$ and together with $C(\Omega)$ if the relations $u_iu_j=e^{\imath B_{i,j}}u_ju_i$ and $u_j f(\omega)u_j^*=f(\tau_j\omega)$ hold. The representation of the unitaries are then the so-called dual magnetic translations. On the algebra $\Aa_d$ are defined unbounded and closed $*$-derivations $\nabla=(\nabla_1,\ldots,\nabla_d)$ by
\begin{equation}
\label{eq-derivrep}
(\nabla_jA)_\omega
\;=\;
\imath [A_{\omega}, X_j]
\;.
\end{equation}
Note that the r.h.s. is again formally covariant, but it is not in $\Aa_d$ for all covariant operator families. The common domain is denoted by $C^1 (\Aa_d)$. On it, the Leibniz rule holds
$$
\nabla (AB)\;=\; (\nabla A)B+A(\nabla B)
\;.
$$
Finally, a positive trace $\Tt$ on $\Aa_d$ is introduced by
\begin{equation}
\label{eq-tracepervol}
 \Tt(A) \; = \;
  \int_{\Omega}
    \PM(d\omega)\;
    \TR\bigl(\langle 0|A_\omega |0\rangle\bigr)
\;.
\end{equation}
Indeed, one has $\Tt(A^*A)\geq 0$ and $\Tt(A^*)=\overline{\Tt(A)}$ as well as $\Tt(AB)=\Tt(BA)$. Furthermore $\Tt(|A B|)\leq \|A\|\,\Tt(|B|)$ and $\Tt(\nabla A)=0$ for $A\in C^1(\Aa_d)$, so that also the partial integration $\Tt(A\nabla B)=-\Tt(\nabla A \,B)$ holds for $A,B\in C^1(\Aa_d)$. Moreover, Birkhoff's ergodic theorem implies that 
\begin{equation}
\label{eq-tracepervol2}
 \Tt(A) \; = \;
      \lim_{N\rightarrow\infty}
       \frac{1}{(2N+1)^d}
        \sum_{|n|\leq N}
         \TR\bigl(\langle n| A_{\omega} |n\rangle\bigr)
\mbox{ , }
\end{equation}
\noindent for $\PM$-almost all $\omega\in\Omega$. This shows that $\Tt$ is the trace
per unit volume.

\vspace{.2cm}

For the description of operators on the half-space Hilbert space $\ell^2(\ZM^{d-1}\times\NM)\otimes\CM^L$, we use as in \cite{KRS,PS} the Toeplitz extension $T(\Aa_d)$ of $\Aa_d$ given by operator families $\widehat{A}=(\widehat{A}_\omega)_{\omega\in\Omega}$ of the form
\begin{equation}
\label{eq-HalfSpaceOps}
\widehat{A}_\omega
\;=\;
\Pi\, A_\omega\,\Pi^*\;+\;\widetilde{A}_\omega
\;,
\end{equation}
where $\Pi:\ell^2(\ZM^d)\otimes\CM^L\to \ell^2(\ZM^{d-1}\times\NM)\otimes\CM^L$ is the partial isometry given by restriction of wave functions and $\widetilde{A}_\omega$ is norm limit of operators acting only on a strip $\ZM^{d-1}\times\{1,\ldots,N\}$ and being covariant in the infinite directions, that is, $V_j\widetilde{A}_\omega V_j^*=\widetilde{A}_{\tau_j\omega}$ for $j=1,\ldots,d-1$. In the terminology of \cite{KRS,PS}, $\widetilde{A}$ lies in the ideal $\Ee_d\subset T(\Aa_d)$ of edge operators. The three C$^*$-algebras form a short exact sequence 
\begin{equation}
\label{eq-ShortExact}
0 \;\rightarrow \;\Ee_d \;\overset{i}{\longrightarrow}\; T(\Aa_d)\;\overset{\pi}{\longrightarrow} \;\Aa_d \;\rightarrow \;0 
\end{equation}
which is used to prove the bulk-boundary correspondence. An alternative abstract way to define $T(\Aa_d)$ is the following (see \cite{PS}).  It is generated by unitaries $\hat{u}_1,\ldots,\hat{u}_{d-1}$, a partial isometry $\hat{u}_d$, $C(\Omega)$ and a supplementary projection $\hat{e}$ such that, on top of the relations $\hat{u}_i\hat{u}_j=e^{\imath B_{i,j}}\hat{u}_j\hat{u}_i$ and $\hat{u}_j f(\omega)\hat{u}_j^*=f(\tau_j\omega)$, one has $\hat{u}_d\hat{u}_d^*=\one-\hat{e}$ and $\hat{u}_d^*\hat{u}_d=\one$, and $\hat{e}$ commutes with $\hat{u}_1,\ldots,\hat{u}_{d-1}$ and $C(\Omega)$. Now let $T_-(\Aa_d)$ be defined similarly from $\breve{u}_1,\ldots,\breve{u}_d$, $C(\Omega)$ and $\breve{e}$, but using the modification $\breve{u}_d\breve{u}_d^*=\one$ and $\breve{u}_d^*\breve{u}_d=\one-\breve{e}$. An isomorphism $\varphi:T_-(\Aa_d)\to T(\Aa_d)$ is obtained by $\varphi(\breve{u}_{j})=\hat{u}_{j}^*$, $\varphi(\breve{e})=\hat{e}$ and $\varphi(f)(\omega)=f(r\omega)$ where $r\omega$ is the reflection of the configuration defined by $r\big((\omega_n)_{n\in\ZM^d}\big)=(\omega_{-n-e_d})_{n\in\ZM^d}$. Further, let $\breve{\pi}_\omega$ be the representations of $T_-(\Aa_d)$ on $\ell^2(\ZM^{d-1}\times \NM_-,\CM^L)$, defined just as the representation $\hat{\pi}_\omega$ of $T(\Aa_d)$ in \cite{PS}. Then with the reflection $R$ defined in \eqref{eq-Reflect}, one has $\hat{\pi}_\omega(\varphi(\breve{A}))=R^*\breve{\pi}_{r\omega}(\breve{A})R$. As $V_jR^*=R^*V_j^*$ for $j=1,\ldots,d-1$, one has the covariance relation $V_j^*\breve{\pi}_{\omega}(\breve{A})V_j=\breve{\pi}_{\tau_j\omega}(\breve{A})$, namely with reversed order to the magnetic translations. Hence the operator in \eqref{eq-What} is covariant w.r.t. $V_j\oplus V_j^*$.

\vspace{.2cm}

On $\Ee_d$ exists a densely defined trace $\widetilde{\Tt}$ defined by
$$
\widetilde{\Tt}
(\widetilde{A})
\;=\;
\EE\;\sum_{n_d\geq 0}
\;\Tr\,\langle 0,n_d|\widetilde{A}_\omega|0,n_d\rangle
\;,
$$
where the $0$ on the r.h.s. is the origin in $\ZM^{d-1}$. This trace is the the trace per unit volume along the boundary combined with the usual trace in the direction perpendicular to the boundary. More informations on this trace and $\widetilde{\Tt}$-traceclass operators can be found in \cite{PS}. 


\end{document}